\documentclass[]{aiaa-tc}

 \title{Processing and Mechanical Property Characterization of Aligned Carbon Nanotube Carbon Matrix Nanocomposites}

 \author{
  Itai Y. Stein%
    \thanks{Graduate Student, Department of Mechanical Engineering.}
  ,\ Hanna M. Vincent%
    \thanks{Undergraduate Student, Department of Materials Science and Engineering.}
  ,\ Stephen A. Steiner III%
    \thanks{Postdoctoral Associate, Department of Materials Science and Engineering.}
  ,\ Elena Colombini%
    \thanks{Visiting Graduate Student, Department of Aeronautics and Astronautics.} ,\\
  and Brian L. Wardle%
    \thanks{Associate Professor, Department of Aeronautics and Astronautics, AIAA Associate Fellow.}\\
  {\normalsize\itshape
   Massachusetts Institute of Technology, Cambridge, MA, 02139}\\
 }

 \AIAApapernumber{YEAR-NUMBER}
 \AIAAconference{Conference Name, Date, and Location}
 \AIAAcopyright{\AIAAcopyrightD{YEAR}}

 \usepackage{wrapfig}
 \usepackage{threeparttable}
 \usepackage{dcolumn}
  \newcolumntype{d}{D{.}{.}{-1}}
 \usepackage{nomencl}
  \makeglossary
 \usepackage{subfigure}
 \usepackage{subfigmat}
 \usepackage{fancyvrb}
  \fvset{fontsize=\footnotesize,xleftmargin=2em}
 \usepackage{lettrine}
\usepackage[colorlinks]{hyperref}
\hypersetup{pdfborder={0 0 0},
	colorlinks=true,
	linkcolor=black,
	citecolor=blue,
	urlcolor=blue
	}
\begin{document}

\maketitle

\begin{abstract}
Materials comprising carbon nanotube (CNT) aligned nanowire (NW) polymer nanocomposites (A-PNCs) are emerging as next-generation materials for use in aerospace structures. Enhanced operating regimes, such as operating temperatures, motivate the study of CNT aligned NW ceramic matrix nanocomposites (A-CMNCs). Here we report the synthesis of CNT A-CMNCs through the pyrolysis of CNT A-PNC precursors, thereby creating carbon matrix CNT A-CMNCs. Characterization reveals that the fabrication of high strength, high temperature, lightweight next-generation aerospace materials is possible using this method. Additional characterization and modeling are planned.
\end{abstract}

\section*{Nomenclature}

\begin{tabbing}
  XXXXXXXXXXX \= \kill
  $A-CMNC$ \> aligned nanowire ceramic matrix nanocomposite \\
  $A-PNC$ \> aligned nanowire polymer matrix nanocomposite \\
  $CNT$ \> carbon nanotube \\
  $E$ \> elastic modulus \\
  $FRP$ \> fiber reinforced plastic \\
  $NW$ \> nanowire \\
  $PyC$ \> pyrolytic carbon \\
  $SEM$ \> scanning electron microscopy \\
  $k$ \> thermal conductivity \\
 \end{tabbing}

\section{Introduction}

\lettrine[nindent=0pt]{T}{he} exceptional electrical \cite{Bezryadin2000, Mooij2006, Wang2010, Xu2008}, thermal \cite{Shen2010, Zhang2011}, and mechanical properties \cite{Wu2005, Chen2006, Wen2008} of nanowires (NWs), which are high aspect ratio nanostructures with diameters on the order of $1 - 100$ nanometers, makes them ideal for a wide range of high performance applications. Due to their high theoretical intrinsic elastic modulus ($E$) of 1 TPa \cite{Wong1997, Walters1999, Yu2000} (about twice that of SiC and five times that of steel), high theoretical intrinsic thermal conductivity ($k$) of over $1000 \ W/mK$ \cite{Kim2001, Baughman2002} (more than $2$ times higher than that of copper), and potential for superconductivity through ballistic electronic transport \cite{Frank1998, Liang2001, Baughman2002, Kociak2001, Tang2001}, CNTs could enable the design and fabrication of next-generation multifunctional, lightweight, structures with highly anisotropic properties. These architectures could potentially be used simultaneously for heat dissipation and shielding, radiation shielding, and energy harvesting, thereby integrating several functions beyond structural components. By controlling the alignment of carbon nanotubes in the matrix of a composite, the fabrication of a material with physical properties that can be precisely and independently tuned becomes possible. We called this controlled-morphology macroscopic material an aligned-CNT nanocomposite. Most aligned-CNT nanocomposites, which have the potential to far surpass the fiber reinforced plastics (FRPs)  currently used in aerospace applications, are polymer nanocomposites, known as CNT aligned NW polymer matrix nanocomposites (A-PNCs). However, CNT A-PNCs are limited by their relatively low operating temperatures, due to thermal degradation of the matrix, and low $k$, due to phonon scattering in the interfaces between the matrix and the highly conducting carbon nanotubes\cite{Panzer2010, Marconnet2011}. To overcome these possible limitations of CNT A-PNCs in high temperature high hardness applications, we study here the substitution of the polymer matrix with a pyrolytic carbon (PyC) ceramic matrix, which forms a nanocomposite we call a CNT aligned NW ceramic matrix nanocomposite (A-CMNC). We believe CNT A-CMNCs have the potential to meet the challenges of extreme environments, where CNT A-PNCs are no longer thermodynamically stable.

\begin{figure}[t!]
 \centering
 \includegraphics[width=4.5in]{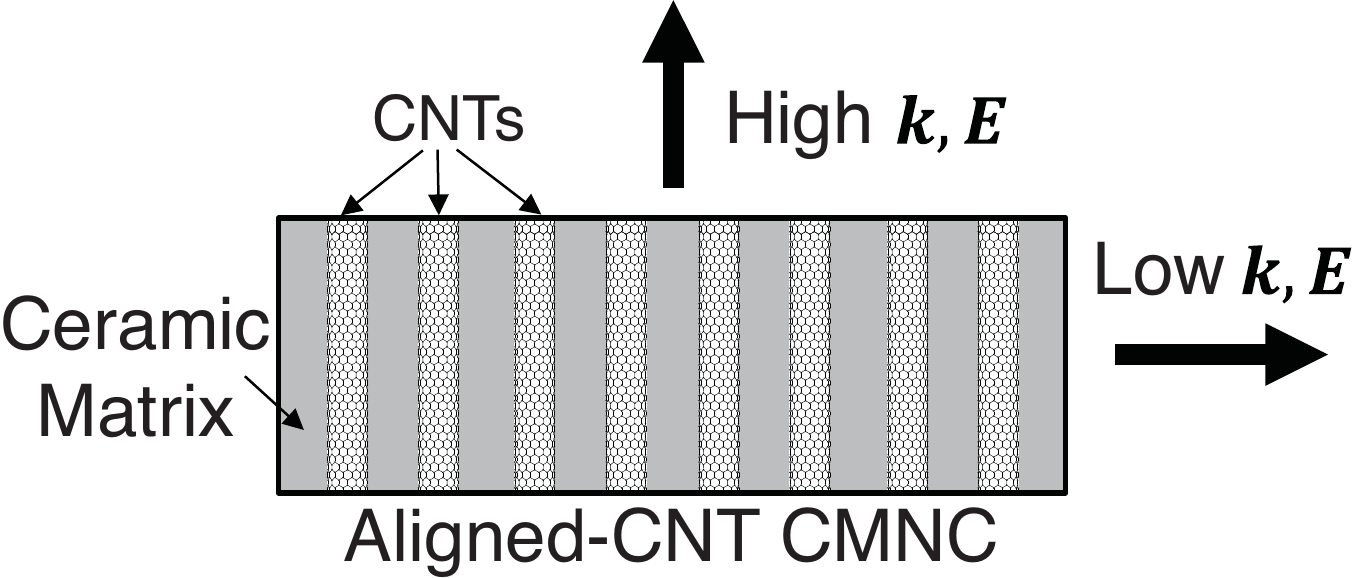}
 \caption{Illustration of the structure of a CNT A-CMNC demonstrating the anisotropy of the physical properties.}
 \label{fig1}
\end{figure}

In this work, CNT A-CMNCs (see Figure~\ref{fig1} for illustration of the architecture) are synthesized via the pyrolysis of CNT A-PNC precursors. The CNT A-PNC precursors are formed by vacuum assisted infusion of vertically aligned multiwalled CNT arrays (also known as forests), grown using thermal chemical vapor deposition\cite{Marconnet2011, Wardle2008, Vaddiraju2009, Liu2011, Garcia2008}, with a commercially available phenolic resin. This fabrication method was previously demonstrated to preserve CNT alignment\cite{Wardle2008}. Using a phenolic resin to synthesize the CNT A-PNC precursors has two advantages: the composites that form are analogous to the epoxy based CNT A-PNCs previously studied in detail\cite{Marconnet2011, Wardle2008, Vaddiraju2009, Liu2011, Cebeci2009, Handlin2013}; the pyrolysis of a phenolic resin composite is a standard process for making high temperature carbon/carbon components\cite{Trick1995}.

To evaluate the performance of CNT A-CMNCs, their surface morphology and mechanical properties were characterized. The surface morphology study, which is primarily concerned with the effect of the pyrolysis process on the alignment of the underlying CNTs, was carried out using scanning electron microscopy (SEM). Using microhardness values collected via Vickers microhardness testing, the mechanical properties of CNT A-CMNCs are quantified and compared to their CNT A-PNC precursors, pure phenolic and pure PyC baseline materials, and four common metal alloys.

\begin{figure}[t!]
 \centering
 \includegraphics[width=6.5in]{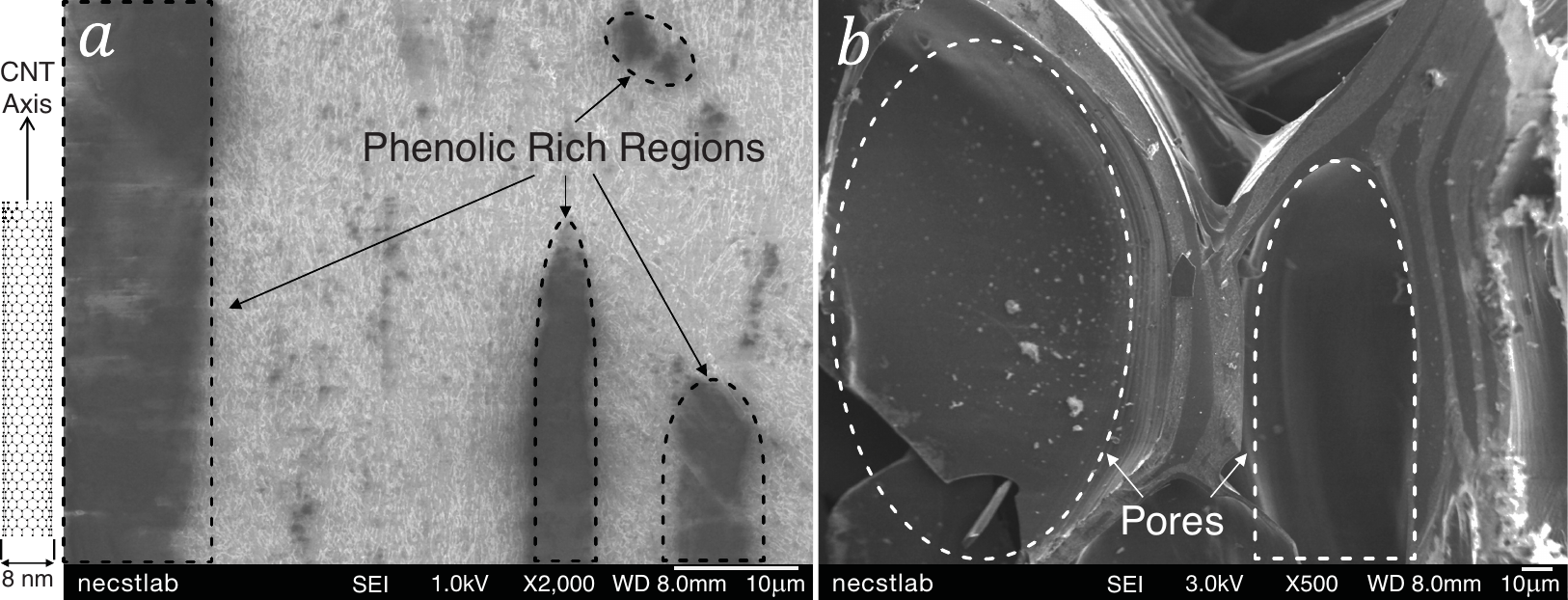}
 \caption{Low resolution SEM micrographs of the cross-section surfaces of fractured CNT A-PNCs (a) and A-CMNCs (b) made using 1 volume \% CNT forests illustrating the phenolic rich regions in the A-PNCs (a) and the A-CMNC porosity that results from pyrolysis.}
 \label{fig2}
\end{figure}

\section{Experimental Methods}

Here we describe how the CNT A-PNCs, and subsequently A-CMNCs, were fabricated, using a synthesized in-house system  of $8\ nm$ diameter CNT forests, and how these nanocomposites were characterized using SEM and Vickers microhardness testing.

\subsection{Nanocomposite Fabrication}

Aligned CNT forests were grown in a $22 \ mm$ internal diameter quartz tube furnace at atmospheric pressure via a previously described thermal catalytic CVD process using ethylene as the carbon source \cite{Marconnet2011, Wardle2008, Vaddiraju2009, Garcia2008}. CNT growth takes place at a nominal temperature of $750 ^\circ C$, with an average growth rate of $\sim 2 \ \mu m/s$ \cite{Marconnet2011, Wardle2008}. The forests were grown on $1\ cm \times 1 \ cm$ Si substrates with a catalytic layer composed of $1 \ nm$ Fe/$10 \ nm \ \mathrm{Al}_{2}\mathrm{O}_{3}$ deposited by electron beam deposition, forming vertically aligned, $\sim 1$ volume $\%$ dense, $8 \ nm$ diameter, millimeter length scale CNT arrays\cite{Marconnet2011, Wardle2008, Vaddiraju2009}. As-grown ($1$ volume $\%$) CNT forests are then delaminated from the Si substrate using a standard lab razor blade, and further processed in their free-standing state.

Fabrication of CNT A-PNCs via vacuum assisted wetting was performed by first gently depositing free-standing CNT forests into a hollow cylindrical mold, ensuring that the primary axis of the CNTs in the forest was orthogonal to the plane of the mold. Phenolic resin was then added on top of the CNT forest, which maintained its orthogonal alignment, but formed some phenolic rich areas (see Figure~\ref{fig2}a). Phenolic infusion took place at $40^\circ$C under vacuum for $\sim24$ hours. CNT A-PNCs were then cured for 6 hours at $80^\circ$C. See Figure~\ref{fig2}a for an image of the cross-sectional surface of a fractured CNT A-PNC.

To convert the cured CNT A-PNCs into A-CMNCs, samples were pyrolyzed as follows in a He environment: $400^\circ$C for 30 minutes, $600^\circ$ for 30 minutes, $750^\circ$ for 30 min. These temperatures are very close to the previously reported temperatures of maximum reaction rate for a neat phenolic resin undergoing pyrolysis\cite{Trick1995}. See Figure~\ref{fig2}b for an image of the cross-sectional surface of a fractured CNT A-CMNC, showing some of the pores that form during the conversion of the phenolic resin into PyC. To quantify the porosity of the PyC matrix, the apparent density of the carbon ceramic samples (both baselines and CNT composites) were approximated by first drying the samples in a He environment, and measuring their mass on a Mettler AE100 Analytical Balance.

\subsection{SEM and Vickers Microhardness Testing}

\begin{figure}[t!]
 \centering
 \includegraphics[width=6.5in]{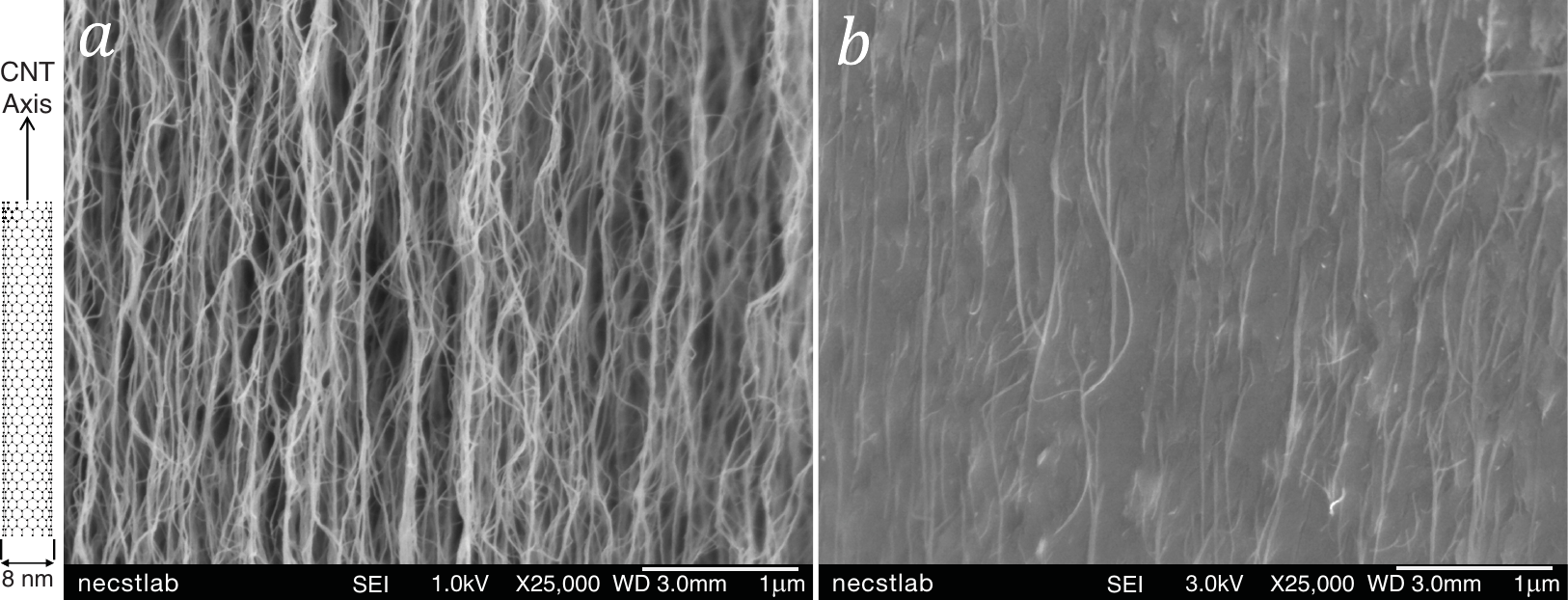}
 \caption{High resolution SEM micrographs for a CNT forest (a) and A-CMNC (b) with a volume fraction of 1 volume \% CNT. These images show that CNT alignment is largely preserved during pyrolysis, but as shown in Figure~\ref{fig2}, some deflection of the CNTs occurs due to pore formation during pyrolysis.}
 \label{fig3}
\end{figure}

\begin{figure}[t!]
 \centering
 \includegraphics[height=2.2in]{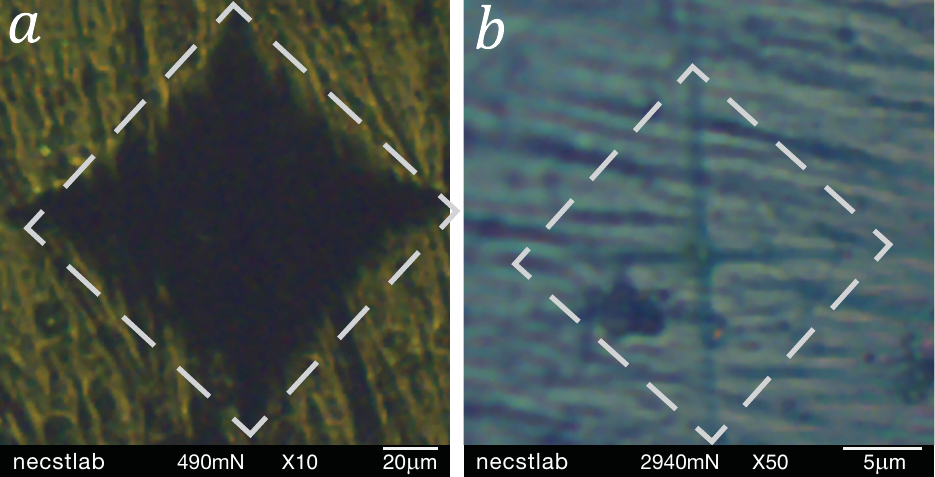}
 \caption{Indentation surfaces for both CNT A-PNCs (a) and A-CMNCs (b). The size of the indentations was used to approximate the Vickers microhardness of value of each sample.}
 \label{fig4}
\end{figure}

To generate representative micrographs of the surface morphology of CNT A-PNCs and A-CMNCs, SEM analysis was performed using a JEOL 6700 cold field-emission gun scanning electron microscope using secondary electron imaging at an accelerating voltage ranging from $1.0$ to $3.0\ kV$ and a working distance ranging from $3.0$  to $8.0\ mm$. To examine whether the pyrolysis step affects the alignment of the CNTs in the A-CMNCs, a cross-sectional fracture surface of A-CMNCs with CNT volume fractions of 1 volume \% was imaged (see Figure~\ref{fig3}b) and compared to the cross-sectional surface of an as-grown CNT forest (see Figure~\ref{fig3}a).

Vickers microhardess numbers for baseline (pure phenolic and PyC), and composite (CNT A-PNCs and A-CMNCs) samples were obtained using a LECO LM 248AT Microhardness tester with a PAXcam camera at $10\times$ to $50\times$ magnification. Once focused on the surface of the samples, the following loads were used to determine the Vickers microhardness: $490\ mN$ for polymer samples (phenolic baselines and CNT A-PNCs); $2940\ mN$ for carbon samples (PyC baselines and CNT A-CMNCs). All measurements were made using ASTM standard C1327-08\cite{ASTM_C1327_08} with a dwell time of 10 seconds, where the final microhardness value reported per samples was the average of 10 indentations. Microhardness values per indentation were approximated using the LECO ConfiDent hardness testing program (version 2.5.2). See Figure~\ref{fig4} for representative Vickers indentation surfaces of CNT A-PNCs (Figure~\ref{fig4}a) and A-CMNCs (Figure~\ref{fig4}b).

\section{Results and Discussion}

As the high resolution SEM micrographs in Figure~\ref{fig3} demonstrate, the alignment of CNTs in A-CMNCs (Figure~\ref{fig3}b) locally does not vary significantly from the CNT alignment in the forests (Figure~\ref{fig3}a). However, while Figure~\ref{fig2}a shows that there is minimal porosity present in the CNT A-PNCs, Figure~\ref{fig2}b clearly illustrates that a significant amount of pores form during the pyrolysis process, which deflect some CNTs in the matrix of the CNT A-CMNCs over a large characteristic length scale. Future work will include additional characterization of the PyC matrix using Brunauer-Emmett-Teller (BET) surface area measurement, to quantify the pore size distribution, and wide angle x-ray scattering (WAXS) to quantify the structure and morphology of the graphitic crystallites that make up the resulting PyC.

\begin{figure}[t!]
 \centering
 \includegraphics[width=6.5in]{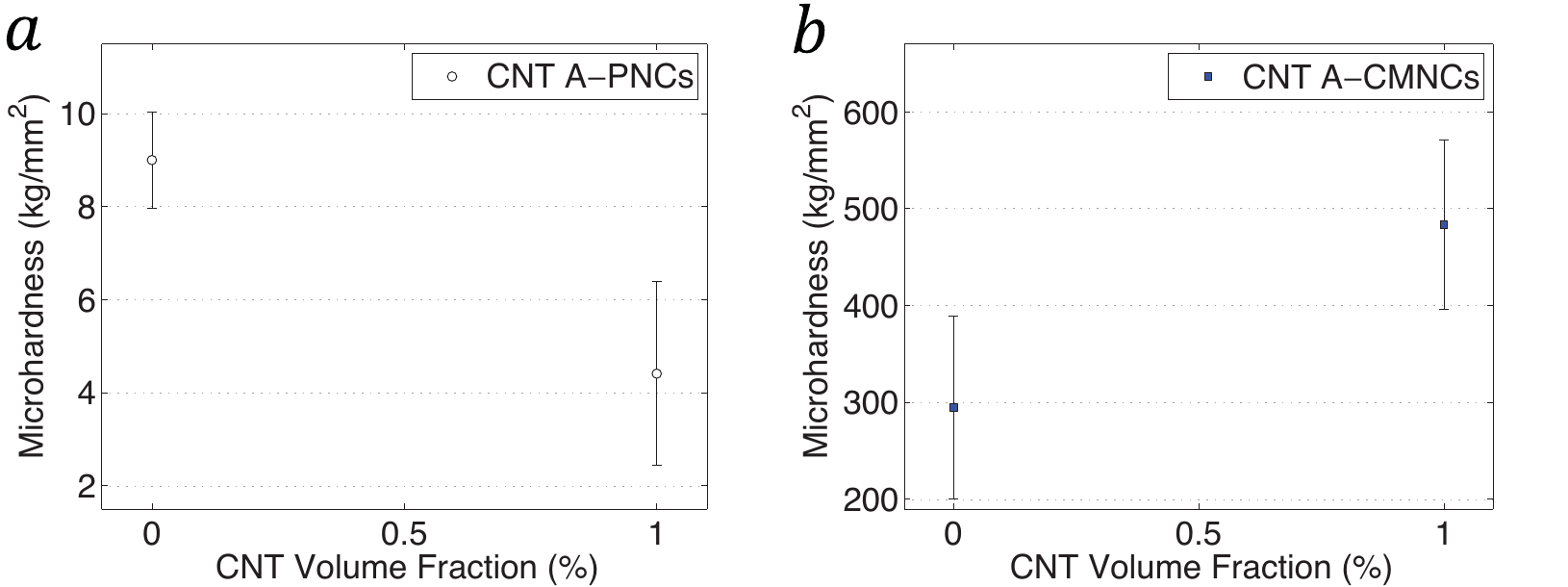}
 \caption{Plot of the experimentally determined Vickers microhardness values for A-PNCs (a) and A-CMNCs (b) at CNT loadings ranging from 0 (baselines) to 1 volume \% showing that pyrolysis leads to more than an order of magnitude increase in the microhardness values of the nanocomposites, and that the presence of CNTs enhances the microhardness of the pyrolyzed composite.}
 \label{fig5}
\end{figure}

To quantify the effect of pyrolysis on the mechanical properties of the composites, the microhardness data collected using Vickers microhardness testing was analyzed. A plot of the Vickers microhardness data can be found in Figure~\ref{fig5}a (polymer samples) and Figure~\ref{fig5}b (PyC samples). As presented in Figure~\ref{fig5}a, the pure phenolic resin baselines (no CNTs present) have a Vickers microhardness value of $9.00 \pm 1.03\ kg/mm^{2}$, while the CNT A-PNCs have a lower Vickers microhardness value of $4.41 \pm 1.97\ kg/mm^{2}$. While a reduction in microhardness following the addition of CNTs may appear counter-intuitive, it is likely a result of the following two factors: poor adhesion of the phenolic resin to the CNTs; the inter-CNT spacing, which is around $78\ nm$ in this case\cite{Stein2013}, is likely on the order of the radius of gyration of the neat phenolic resin ($\sim 100\ nm$) hindering some of the crosslinking of the polymer matrix during curing. Therefore, the hardness of the CNT A-PNCs is most likely reduced by these poor CNT-phenolic interfacial properties, and a lower crosslink density in the phenolic matrix. Such a reduction was previously observed when the effect of CNT thermal boundary resistance on the thermal conductivity of CNT A-PNCs was studied\cite{Duong2009}, demonstrating that the addition of nanoparticles with exceptional intrinsic properties does not guarantee the enhancement of the material properties of the macroscopic architecture. Another important consideration is that, as shown in Figure~\ref{fig2}a, the phenolic infusion may lead to some distortion of the CNT alignment, and could cause resin rich regions on the scale of the indentor to form. Coupled with the fact that the Vickers microhardness test assumes that the materials tested are isotropic, which is not true of the CNT composites, these resin rich regions may lead to microhardness results that are not fully representative of the mechanical properties of the CNT composites. As presented in Figure~\ref{fig5}b, the pure PyC baselines (no CNTs present) have a Vickers microhardness value of $294.7 \pm 94.0\ kg/mm^{2}$, while the CNT A-CMNCs have an enhanced Vickers microhardness value of $483.2 \pm 87.5\ kg/mm^{2}$. As illustrated in Table~\ref{tbl1}, the A-CMNCs have a slightly lower apparent density ($888 \pm 115\ kg/m^{3}$) than the pure PyC baselines ($930 \pm 56\ kg/m^{3}$). This means that the CNT A-CMNCs have a much much higher specific hardness ($\simeq 5.34\ MPa \cdot m^{3}/kg$ for the A-CMNCs vs. $\simeq 3.11\ MPa \cdot m^{3}/kg$ for the PyC matrix), defined as the ratio of the average microhardness, in $MPa$, and apparent density, in $kg/m^{3}$ (see Table~\ref{tbl1}). A rough estimate of the porosity of the PyC matrix of the CNT A-CMNCs can be attained by correcting the apparent density for the mass contributed by the CNTs (which have an intrinsic density of $\approx 1700\ kg/m^{3}$), and dividing by the theoretical density of the graphitic crystallites that make up the PyC matrix (which have an intrinsic density of $\approx 2240\ kg/m^{3}$), thereby yielding the approximate volume fraction of carbon in the matrix of the CNT A-CMNCs. This rough estimation indicates that the matrix of the CNT A-CMNCs has about $60\%$ porosity, illustrating that multiple phenolic infusions and pyrolyzations would be necessary to minimize the porosity of the PyC matrix. Porosity of this magnitude is observed in the low resolution SEM micrographs of the A-CMNCs (see Figure~\ref{fig2}b)

\begin{table}[h!]
 \begin{center}
  \caption{Hardness, apparent density, and specific hardness in SI units for the A-CMNCs, PyC matrix, and the four engineering metal alloys}
  \label{tbl1}
  \begin{tabular}{cccc}
       \hline
       Material & Hardness ($MPa$) & Apparent Density ($kg/m^{3}$) & Specific Hardness ($MPa \cdot m^{3}/kg$)\\
       \hline
       CNT A-CMNC & $4740 \pm 845$ & $888 \pm 115$ & 5.34 \\
       PyC matrix & $2890 \pm 922$ & $930 \pm 56$ & 3.11 \\
       Ti alloy & 2990 & 4420 & 0.68 \\
       Al alloy & 1320 & 2700 & 0.49 \\
       Low carbon steel & 2900 & 7900 & 0.37 \\
       Brass & 1570 & 8430 & 0.19 \\
       \hline
  \end{tabular}
 \end{center}
\end{table}

To put the properties of the CNT A-CMNCs into perspective, four different metal alloys are used as a comparison\cite{Jacobs1992}: a low carbon CrNi-steel (X5 CrNi 189), a Ti alloy (TiAl6V4), an Al alloy (AlMgSi1), and a brass (CuZn40Pb2). The metal alloys have the following microhardness\cite{Jacobs1992} and density values (from their respective data sheet): $296\ kg/mm^{2}$ and $7900\ kg/m^{3}$ for the low carbon steel; $305\ kg/mm^{2}$ and $4420\ kg/m^{3}$ for the Ti alloy; $135\ kg/mm^{2}$ and $2700\ kg/m^{3}$ for the Al alloy; $160\ kg/mm^{2}$ and $8430\ kg/m^{3}$ for the brass. As shown in Table~\ref{tbl1}, the specific hardness of the CNT A-CMNC ($\simeq 5.34\ MPa \cdot m^{3}/kg$) far exceeds all the values calculated for the four different metal alloys: $\simeq 0.37\ MPa \cdot m^{3}/kg$ for the low carbon steel; $\simeq 0.68\ MPa \cdot m^{3}/kg$ for the Ti alloy; $\simeq 0.49\ MPa \cdot m^{3}/kg$ for the Al alloy; $\simeq 0.19\ MPa \cdot m^{3}/kg$ for the brass. Therefore, the CNT A-CMNCs reported here outperform a variety of common engineering metal alloys, while having a much lower density, potentially enabling a large degree of weight savings in next-generation aerospace architectures, while retaining extreme environment operating capabilities.

\section{Conclusion}

In conclusion, the fabrication process of CNT A-CMNCs, which show microhardness enhancements when compared to pure PyC samples, through the pyrolysis of CNT A-PNC precursors was demonstrated. Also, SEM analysis illustrates that the CNT alignment is preserved in the CNT A-CMNCs. Future work will include: a study on whether the porosity of the PyC matrix of the CNT A-CMNCs can be reduced through multiple phenolic infiltrations and pyrolyzations; the quantification of the effect of CNT volume fraction on the enhancement of the Vickers microhardness of CNT A-CMNCs. While SEM analysis showed that there was significant porosity present in the CNT A-CMNCs, their effect on the mechanical properties, was not yet quantified, and will be studied both experimentally and theoretically in the future. Finally, since the effect of the pyrolyzation temperature on the intrinsic properties of the CNTs is fundamental to the properties of their composites, but was not studied here, further study is necessary to evaluate the quality of the CNTs post pyrolyzation.

\section*{Acknowledgments}

The authors would like to thank Sunny Wicks (MIT), Samuel Buschhorn (MIT), and Dr. Noa Lachman-Senesh (MIT) for helpful discussions, and John Kane (MIT) and the entire necstlab at MIT for technical support and advice. This work made use of the core facilities at the Institute for Soldier Nanotechnologies at MIT, supported (in part) by the U.S. Army Research Office under contract W911NF-07-D-0004, and the Undergraduate Teaching Laboratory in the Department of Materials Science and Engineering at MIT.

\end{document}